\documentclass[twocolumn,amsmath,amssymb,prb]{revtex4}
\usepackage{graphicx}
\usepackage{dcolumn}
\usepackage{bm}
\usepackage{color}
\definecolor{blue}{rgb}{0.,.0,1.0}

\newcommand \boldeta {\mbox{\boldmath $\eta$}}

\def \be {\begin{equation}}
\def \ee {\end{equation}}
\def \ben {\begin{eqnarray}}
\def \een {\end{eqnarray}}

\begin{document}

\bibliographystyle{../prsty}

\title{Can quantum transition state theory be defined as an exact $t=0_+$ limit?}

\author{Seogjoo Jang\footnote{Email:sjang@qc.cuny.edu}}
\affiliation{Department of Chemistry and Biochemistry, Queens College, City University of New York, 65-30 Kissena Boulevard, Queens, New York 11367\footnote{mailing address}  \& PhD programs in Chemistry and Physics, and Initiative for the Theoretical Sciences, Graduate Center, City University of New York, 365 Fifth Avenue, New York, NY 10016}
\author{Gregory A. Voth\footnote{Email:gavoth@uchicago.edu}}
\affiliation{Department of Chemistry, James Franck Institute, Institute for Biophysical Dynamics and Computation Institute, University of Chicago, 5735 S. Ellis Avenue, Chicago, Illinois 60637}

\date{Accepted for Publication in the Journal of Chemical Physics, Feb. 9, 2016}

\begin{abstract}
The definition of the classical transition state theory (TST) as a $t\rightarrow 0_+$ limit of the flux-side time correlation function relies on the assumption that simultaneous measurement of population and flux is a well defined physical process.    However, the noncommutativity of the two measurements in quantum mechanics makes the extension of such a concept to the quantum regime impossible.  For this reason, quantum TST (QTST) has been generally accepted as any kind of quantum rate theory reproducing the TST in the classical limit, and there has been a broad consensus that no unique QTST retaining all the properties of TST can be defined.    Contrary to this widely held view, Hele and Althorpe (HA) [J. Chem. Phys. {\bf 138}, 084108 (2013)] recently suggested that a true QTST can be defined as the exact $t\rightarrow 0_+$ limit of a certain kind of quantum flux-side time correlation function and that it is equivalent to the ring polymer molecular dynamics (RPMD) TST.  This work seeks to question and clarify certain  assumptions underlying these suggestions and their implications.  First, the time correlation function used by HA as a starting expression is not related to the kinetic rate constant by virtue of linear response theory, which is the first important step in relating a $t=0_+$ limit to a physically measurable rate.  Second, a theoretical analysis calls into question a key step in HA's proof which appears not to rely on an exact quantum mechanical identity.  The correction of this makes the true $t=0_+$ limit of HA's QTST different from the RPMD-TST rate expression, but rather equal to the well-known path integral quantum transition state theory rate expression for the case of centroid dividing surface.  An alternative quantum rate expression is then formulated starting from the linear response theory and by applying a recently developed formalism of real time dynamics of imaginary time path integrals [S. Jang, A. V. Sinitskiy, and G. A. Voth, J. Chem. Phys. {\bf 140}, 154103 (2014)].  It is shown that the $t\rightarrow 0_+ $ limit of the new rate expression vanishes in the exact quantum limit.      
\end{abstract}


\maketitle

\section{Introduction}
How to extend the transition state theory (TST)\cite{pechukas-r1,hynes-r1,chandler-jsp42,hanggi-rmp62,pollak-r1} to the quantum regime has been a long standing theoretical challenge.\cite{mclafferty-cpl27,miller-acc26}  Within the classical mechanics, the TST is a well defined theory with firmly established computational methods and simulation protocols. However, the very concept of measuring flux at a localized dividing surface is at odds with the quantum mechanical uncertainty principle.  Moreover, the notion of sorting out ``trajectories" that do not recross, which allows the TST to be the upper bound of a true barrier crossing rate, is difficult to envision quantum mechanically.    In fact, it does not seem clear whether a practical quantum TST (QTST) that translates all the assumptions of the TST to quantum regime can be developed at all. 

While genuine QTSTs serving as a rigorous upper bound for quantum barrier crossing rates have been developed,\cite{mclafferty-cpl27,pollak-jcp74,pollak-jcp107} they are either difficult to implement or lack quantitative accuracy.  Thus, here we adopt a loose definition of QTST as a quantum barrier crossing rate theory approaching the TST in the classical limit.  Various QTSTs that allow practical calculations have been developed based on a wide range of theoretical approaches.    

One approach is to invoke approximate time dependent quantum dynamics\cite{voth-jpc93,voth-jcp91,voth-jpc97,hansen-jcp,hansen-jpc,pollak-jcp108,jang-jcp112} near the barrier top, which results in time dependent rates, and to define the QTST as the steady state limit of the time dependent rate expression. Another approach is to abandon explicit consideration of the dynamics from the outset and to extract the rate from the quantum partition function representing metastable reactant states.\cite{affleck-prl46,gillan-jp-c20,messina-jcp98,messina-jcp99,mills-cpl278}  Justification for this latter approach can be made  through analytic continuation of the partition function to the complex time domain or by an argument of detailed balance.  Theoretical formulations unifying the two approaches have also been developed.\cite{cao-jcp105} 

Recently,  Hele and Althorpe (HA) proposed a new formulation of QTST,\cite{hele-jcp138} hereafter termed as HA-QTST,  and suggested that they have proven its equivalence to the ring polymer molecular dynamics TST (RPMD-TST).\cite{craig-jcp122,craig-jcp123}   They argued that HA-QTST is a true QTST in the sense that it corresponds to an exact $t\rightarrow 0_+$ limit of a quantum flux-side correlation function.  A follow-up work\cite{althorpe-jcp139} presented another formal analysis suggesting again that the HA-QTST is exact under a condition of no recrossing, which however was defined in a formal way without a physical definition of quantum mechanical recrossing.  A careful theoretical examination of this overall formalism  issue is therefore important for our general understanding of QTST.

A distinctive aspect of HA-QTST is that the formulation starts from a special kind of quantum time correlation function which is apparently constructed based on a mathematical consideration of its $t\rightarrow 0_+$ limit, the physical basis of which is not clear. In addition, as will be analyzed in detail in Sec. III,  HA's proof for the equivalence between the $t\rightarrow 0_+$ limit of their quantum correlation function and the RPMD-TST\cite{craig-jcp122,craig-jcp123} rate can evidently only be understood through an incorrect or at best approximate application of a quantum mechanical identity.   When this is corrected, we find that the true limit of HA-QTST for the case of centroid dividing surface becomes identical to the well-known path integral quantum transition theory (PI-QTST) expression above the crossover temperature,\cite{voth-jcp91,voth-jpc97} which can be understood in a much simpler manner\cite{voth-jcp91,voth-jpc97,jang-jcp112,geva-book} and amounts to classical approximation for the dynamical factor. Thus, as will be explained in Sec. III, the implications\cite{hele-jcp138,althorpe-jcp139} of HA-QTST need to be reassessed.    

As a formalism alternative to HA-QTST,  we then present a new path integral based quantum rate expression, starting from the well established Yamamoto expression\cite{yamamoto-jcp33,voth-jpc93,stuchebrukhov-jcp95} for the exact quantum rate based on the linear response theory,\cite{kubo-jpsj-12,kubo-jpsj-12-2} and employing a recently developed quantum dynamics formalism,\cite{jang-cmd-rpmd,jang-jcp140} hereafter called as real time dynamics of  imaginary time path integral (RDIP).   In the classical limit, the $t=0_+$ limit of our rate expression becomes equal to classical TST.  However, if the quantum limit is taken first, the $t=0_+$ limit of the rate expression can be shown to be zero due to quantum delocalization.  This is consistent with other work,\cite{costley-cpl83} and confirms the assessment\cite{miller-acc26} that constructing a rigorous QTST as the $t=0_+$ limit of a quantum dynamics is likely to be impossible.  

The remainder of this paper is organized as follows.  Section II provides a review of the linear response theory as a preliminary step.  Section III presents a detailed analysis of HA-QTST.  
Then, in Sec. IV, we develop a new formulation of QTST employing the RDIP approach.  Sec. V provides concluding remarks.  

\section{Time dependent rate expression based on quantum linear response theory}
\begin{figure}
\includegraphics[width=3in]{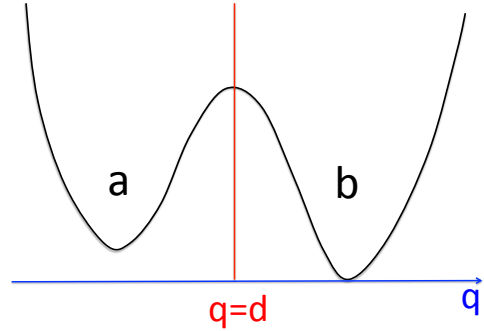}
\caption{A schematic of free energy profile along the reaction coordinate, $q$.  The reactant is denoted as $a$ and the product is denoted as $b$.  $q=d$ is the dividing surface between the two.} 
\end{figure}

As a preliminary step, we here provide a short review of the exact quantum rate expression\cite{yamamoto-jcp33,voth-jpc93,stuchebrukhov-jcp95} based on the linear response theory.\cite{kubo-jpsj-12,kubo-jpsj-12-2}This will serve as the basis for the discussion in Sec. III and the formulation developed in Sec. IV.    

Let us consider an activated rate process from a reactant ($a$) to a product  ($b$). Figure 1 shows a schematic of the free energy profile.  Here, we assume that the reaction occurs in a one dimensional coordinate $q$, but the formalism can be readily extended to multidimensional situation.  The population functions of $a$ and $b$, denoted here as $h_a(q)$ and $h_b(q)$, respectively, satisfy the condition of $h_{a} (q) +h_{b} (q)=1$.
We here make the simplest and well known choice of  $h_b(q)=\Theta (q-d)$, where $\Theta$ is the step function and $d$ is the dividing surface (point for one dimensional case) between $a$ and $b$.    The quantum mechanical population operators are then defined as $\hat h_a=h_a(\hat q)$ and $\hat h_b= h_b(\hat q)$, and the flux operator from $a$ to $b$ can be expressed as
\be
\hat F=\frac{i}{\hbar} [\hat H,h_b (\hat q)]=\frac{\hat p}{2m} h_b'({\hat q})+h_b'({\hat q})\frac{\hat p}{2m}\ , \label{eq:flux_op}
\ee
where $\hat p$ is the momentum operator conjugate to $\hat q$, $m$ is the mass of the particle, and $h_b'(q)=\delta (q-d)$.   Starting from the linear response theory\cite{kubo-jpsj-12,kubo-jpsj-12-2} and the assumption of rate behavior in which the longest time scale of the system is the rate of interconversion from wells $a$ to $b$, as detailed in Appendix A, on can obtain the following expression for the time dependent forward rate constant from $a$ to $b$:
\be
k(t)=\frac{1}{Z_a} \frac{1}{\beta} \int_0^\beta d\lambda\ Tr \{ e^{-(\beta-\lambda)\hat H} \hat h_a e^{-\lambda \hat H} \hat F(t)\} \ , \label{eq:kt-sf}
\ee
where $\beta=1/(k_BT)$ and $Z_a=Tr\left \{e^{-\beta \hat H}\hat h_a \right\}$, the reactant partition function.  The operator $\hat F(t)$ in the above expression is the time dependent flux operator defined as 
\ben
\hat F(t)&=&e^{i\hat Ht/\hbar}\hat F e^{-i\hat Ht/\hbar}\nonumber \\
&=&\frac{\hat p(t)}{2m} h_b'({\hat q}(t))+h_b'({\hat q}(t))\frac{\hat p(t)}{2m}\ , \label{eq:Ft_def}
\een 
where $\hat p(t)=e^{i\hat Ht/\hbar}\hat p e^{-i\hat H t/\hbar}$ and $\hat q(t)=e^{i\hat Ht/\hbar} \hat q e^{-i\hat H t/\hbar}$.  Utilizing the fact that  $\hat F(t)=i[\hat H, \hat h_b(t)]/\hbar$ and the cyclic invariance of the trace operation, 
we can rewrite Eq. (\ref{eq:kt-sf}) as follows:
\ben
k(t)&=&-\frac{1}{Z_a} \frac{1}{\beta} \int_0^\beta d\lambda Tr \{ e^{-(\beta-\lambda)\hat H} \frac{i}{\hbar}[\hat H,\hat h_a] e^{-\lambda \hat H} \hat h_b(\hat q(t))\}\nonumber \\
&=&\frac{1}{Z_a} \frac{1}{\beta} \int_0^\beta d\lambda Tr \{ e^{-(\beta-\lambda)\hat H} \hat F e^{-\lambda \hat H} \hat h_b(\hat q(t))\} \ ,\label{eq:kt-fs}
\een
where, in the second equality, the fact that $\hat F=-i[\hat H,\hat h_a]/\hbar$ has been used.
For direct quantum dynamics simulation, Eq. (\ref{eq:kt-fs}) may be more advantageous than Eq. (\ref{eq:kt-sf})  because it allows sampling of trajectories starting only from the barrier region. However, the evaluation of the exact quantum flux operator still remains a challenging task. 

It can be shown that the exact $t=0_+$ limit of Eq. (\ref{eq:kt-sf}) or (\ref{eq:kt-fs}) is zero, which is a manifestation of fundamental quantum principles.  Since the population and flux operators do not commute, measurement of the flux right after the preparation of the system, {\it i.e.}, population measurement, results in both positive and negative fluxes with an equal probability.  Thus, the structure of the linear response theory dictates that the exact $t=0_+$ limit of the reactive flux is zero in the quantum regime.  Therefore, any quantum rate formulation based on the exact $t=0_+$ limit should address how this fundamental quantum mechanical property can be somehow avoided. 
 
\section{QTST by Hele and Althorpe}
One feature that sets the HA-QTST\cite{hele-jcp138} apart from other earlier theories is that,  instead of Eq. (\ref{eq:kt-sf}) or (\ref{eq:kt-fs}), the rate formulation starts from  a quantum time correlation function they have termed as generalized Kubo transform of the side-side correlation function (GKSCF).  

The apparent motivation\cite{hele-jcp138} for introducing GKSCF is to identify a time correlation function with nonzero $t=0_+$ limit, and HA has constructed it starting from a complex time path integral representation of the quantum partition function $Z=Tr\{e^{-\beta \hat H}\}$. In our notation, as detailed in Appendix B,  the GKSCF can be expressed as 
\ben
&&\tilde C_{ss}(t) = \int dq_{_1}\cdots \int dq_{_P}\int d\eta_{_1}\cdots \int d\eta_{_P}\nonumber \\
&&\times \int dz_1\cdots \int dz_{_P} \rho({\bf q},\boldeta){\mathcal G}({\bf q}, \boldeta,{\bf z};t)   h_b\left (f({\bf q})\right) h_b \left (f({\bf z})\right)   \ ,  \nonumber \\\label{eq:c_sst}
\een
where ${\bf q}$ and $\boldeta$ are path coordinates defined by Eqs. (\ref{eq:qi_def}) and (\ref{eq:etai_def}),  $f({\bf q})$ is a symmetric function of $q_k$'s defined in the space of imaginary time paths. The density $\rho({\bf q},\boldeta)$ is defined by Eq. (\ref{eq:def-rho}) and represents the canonical density operator.  The term ${\mathcal G}({\bf q},\boldeta,{\bf z};t)$ is defined by  Eq. (\ref{eq:def-g}) and retains all the dynamical information.  Note that we have incorporated the position of the dividing surface $d$ into the definition of the population function, unlike the original formulation by HA.\cite{hele-jcp138}  Thus, $f({\bf q})$ in Eq. (\ref{eq:c_sst}) is independent of $d$.  

Although mathematically well defined, the physical meaning of GKSCF is less clear.  HA rendered the $t=0_+$ limit of the time derivative of the correlation function to be non-zero by making the flux and side dividing surfaces equal.\cite{hele-jcp138}  However, this procedure is not possible if one adheres to the rules of quantum mechanics and the requirement to express the rate in an exact and general form, {\it e.g.}, from the linear response theory [Eqs. (\ref{eq:kt-sf})-(\ref{eq:kt-fs})].   This is also illustrated by the fact that Fig. 2 of Ref. \onlinecite{hele-jcp138} depicts the manipulation of the Feynman diagrams for the correlation function, which appear to be disallowed due to the non-commutation of the operators.

It is possible to further analyze the GKSCF as follows.
From the definition of ${\mathcal G}({\bf q}, \boldeta,{\bf z};t)$ given by Eq. (\ref{eq:def-g}), it is clear that 
\be
{\mathcal G}({\bf q}, \boldeta,{\bf z};0)=\delta ({\bf q}-{\bf z})\delta (\boldeta) \ . \label{eq:gqz0}
\ee  
Thus, for $t=0$, Eq. (\ref{eq:c_sst}) becomes
\be
\tilde C_{ss}(0) = \int dq_{_1}\cdots \int dq_{_P}\ \rho({\bf q},0) h_b\left (f({\bf q})\right)   \ , \label{eq:css0}
\ee
where  $\rho({\bf q},0)=\prod_{k=1}^P \langle q_k|e^{-\epsilon \hat H}|q_{k+1}\rangle$, with $\epsilon=\beta/P$ and the cyclic boundary condition of $q_{_{P+1}}=q_1$.   
Note that the population function $h_b$ is a nonlinear function.  As a result, Eq. (\ref{eq:css0}) does not have any corresponding quantum mechanical operator expression.   This is because the imaginary time in the Feynman path integral is a fictitious time, each labeling a different realization of the system in the quantum canonical ensemble. Any function that  depends nonlinearly on the collection of the imaginary time path coordinates mixes up different realizations of the ensemble in a nonlinear manner.  Such a function has temperature as an implicit variable, and does not have any analogue for the case of a pure quantum state.  Thus,  $h_b\left (f({\bf q})\right)$ cannot be related to any genuine physical observable except in the classical limit or zero temperature limit.  This is true even for the case where $f({\bf q})=q_0=(q_1+\cdots+q_{_p})/P$, namely the imaginary time path centroid.  Therefore, it is not possible to establish unambiguous relationship between $\tilde C_{ss}(0)$ (and thus $\tilde C_{ss}(t)$) with physically measurable quantities.  This is in strong contrast with the physical basis for Eq. (\ref{eq:kt-sf}) or (\ref{eq:kt-fs}), as demonstrated in detail in Appendix A.  Thus, it appears that the starting expression used by HA\cite{hele-jcp138} lacks a clear physical justification unlike the linear response rate expression.\cite{yamamoto-jcp33,voth-jpc93,stuchebrukhov-jcp95}  This does not necessarily mean that the time derivative of $\tilde C_{ss}(t)$ cannot be used as a quantum rate.  However, no fundamental physical basis appears to exist which suggests that such a limit should be considered as a true QTST.\cite{hele-jcp138}  With this issue clarified, we can now provide a further analysis of HA-QTST to discuss another important issue.

As the next step, HA calculates the time derivative of  Eq. (\ref{eq:c_sst}) as follows:
\ben
&&\tilde C_{fs}(t)\equiv -\frac{d}{dt}\tilde C_{ss}(t) \nonumber \\
&&= -\int dq_{_1}\cdots \int dq_{_P}\int d\eta_{_1}\cdots \int d\eta_{_P}\rho({\bf q},\boldeta)\nonumber \\
&&\times \int dz_1\cdots \int dz_{_P} h_b(f({\bf q}))h_b(f({\bf z})) \frac{d}{dt}{\mathcal G}({\bf q}, \boldeta,{\bf z};t)  \nonumber\ , \\     \label{eq:dc_sst}
\een
where the detailed expression for $d{\mathcal G}({\bf q}, \boldeta, {\bf z};t)/dt$ is given by Eq. (\ref{eq:dgdt}) in Appendix C. 
Through partial integration of the resulting expression for Eq. (\ref{eq:dc_sst}), as detailed in Appendix C and using the definition that $h_b$ is a step function, we obtain the following expression
\ben
&&\tilde C_{fs}(t) = \int dq_{_1}\cdots \int dq_{_P}\int d\eta_{_1}\cdots \int d\eta_{_P}\nonumber \\
&&\times \int dz_1\cdots \int dz_{_P}  \rho({\bf q},\boldeta) {\mathcal G}({\bf q}, \boldeta,{\bf z};t) \nonumber \\
&&\hspace{.5in}\times \Theta(f({\bf z})-d) \delta(f({\bf q})-d)\nonumber \\
&& \times  \frac{1}{2m}\sum_{k=1}^P \frac{\partial f({\bf q})}{\partial q_k} \left \{ \bar p_+(q_{k-1}-\frac{\eta_{k-1}}{2},q_k+\frac{\eta_k}{2};\epsilon)   \right .  \nonumber \\ 
&& \left . \hspace{1in}+\bar p_-(q_k-\frac{\eta_k}{2},q_{k+1}+\frac{\eta_{k+1}}{2};\epsilon) \right \} \ , \nonumber \\ \label{eq:dc_sst-1}
\een
where we have introduced new imaginary time momentum averages as follows:  
\ben
\bar p_+(x',x'';\epsilon)= \frac{\langle x'|e^{-\epsilon\hat H}\hat p|x''\rangle}{\langle x'|e^{-\epsilon\hat H}|x''\rangle}\ , \label{eq:p+}\\
\bar p_-(x',x'';\epsilon)=\frac{\langle x'|\hat p e^{-\epsilon \hat H} |x''\rangle }{\langle x'|e^{-\epsilon \hat H} |x''\rangle}\ . \label{eq:p-}
\een
Equation (\ref{eq:dc_sst-1}) can be shown to be equivalent to Eq. (31) of Ref. \onlinecite{hele-jcp138}.   At $t=0$, it becomes 
\ben
&&\hspace{-.2in}\tilde C_{fs}(0) = \int dq_{_1}\cdots \int dq_{_P} \rho({\bf q},0) \Theta(f({\bf q})-d) \delta(f({\bf q})-d)\nonumber \\ 
&& \times  \frac{1}{2m}\sum_{k=1}^P \frac{\partial f({\bf q})}{\partial q_k} \left \{ \bar p_+(q_{k-1},q_k;\epsilon) +\bar p_-(q_k,q_{k+1};\epsilon) \right \} \ \ , \nonumber \\
\ \label{eq:dc_sst-10}\ 
\een
where the identity of Eq. (\ref{eq:gqz0}) has been used.  This expression can be shown to be zero, as detailed in Appendix D, due to the cyclic symmetry of the imaginary time path integral.  

Invoking an analogy to the classical TST, HA defines their QTST as the $t=0_+$ limit of Eq. (\ref{eq:dc_sst-1}).   For this, they use a short time approximation for real time propagators, $e^{\pm it\hat H/\hbar}$'s, within the definition of ${\mathcal G}({\bf q}, \boldeta, {\bf z};t)$, Eq. (\ref{eq:def-g}). This procedure leads to classical-like real time momenta $p_k$'s satisfying the following relation:\footnote{However, it should be made clear that $p_k$ is simply an integrand in the unit of momentum and that no apparent theoretical justification is available for the suggestion that it is a dynamical variable following the RPMD equation of motion as in the RPMD-TST.\cite{craig-jcp122,craig-jcp123}}   
\be
z_k=q_k+\frac{p_k}{m}t \ .  \label{eq:zk_pk}
\ee
HA then replaces the integrands $z_k$'s in Eq. (\ref{eq:dc_sst-1}) with $p_k$'s, which is mathematically valid as long as $t>0$. Taking the limit of $t=0_+$, the resulting expression can be shown to be
\ben
&&k_{HA}Z_a=\tilde C_{fs}(0_+) \nonumber \\
&&= \frac{1}{(2\pi \hbar)^P}\int dq_{_1}\cdots \int dq_{_P}\int d \eta_{_1}\cdots \int d \eta_{_P}\nonumber \\
&&\hspace{.1in}\times \int dp_1\cdots \int dp_{_P}  \rho({\bf q},\boldeta) \exp (i{\bf p} \cdot \boldeta /\hbar) \nonumber \\
&&\hspace{.1in}\times \Theta (f({\bf q}+\frac{\bf p}{m}0_+)-d) \delta(f({\bf q})-d) \nonumber \\
&& \hspace{.1in} \times  \frac{1}{2m}\sum_{k=1}^P \frac{\partial f({\bf q})}{\partial q_k} \left \{ \bar p_+(q_{k-1}-\frac{\eta_{k-1}}{2},q_k+\frac{\eta_k}{2};\epsilon)   \right .  \nonumber \\ 
&& \left . \hspace{.8in}+\bar p_-(q_k-\frac{\eta_k}{2},q_{k+1}+\frac{\eta_{k+1}}{2};\epsilon) \right \} \ .  \label{eq:dc_sst-10+1}
\een
Based on the following expansion 
\be
f({\bf q}+\frac{\bf p}{m}0_+)=f({\bf q})+\frac{{\bf p}}{m}\cdot \nabla f({\bf q})0_+\ ,
\ee
and employing the constraint of $f({\bf q})=d$, one can make the following replacement: 
\be 
\Theta(f({\bf q}+\frac{\bf p}{m}0_+)-d) =\Theta \left (\bf p\cdot \nabla f({\bf q}) \right)  \ .
\ee
Thus, Eq. (\ref{eq:dc_sst-10+1}) can be expressed as
\ben
&&k_{HA}Z_a= \frac{1}{(2\pi \hbar)^P}\int dq_{_1}\cdots \int dq_{_P}\int d \eta_{_1}\cdots \int d \eta_{_P}\nonumber \\
&&\hspace{.2in}\times \int dp_1\cdots \int dp_{_P}  \rho({\bf q},\boldeta) \exp (i{\bf p} \cdot \boldeta /\hbar) \nonumber \\
&&\hspace{.2in}\times   \Theta \left ({\bf p}\cdot \nabla f({\bf q}) \right) \delta (f ({\bf q})-d) \nonumber \\
&& \hspace{.2in} \times  \frac{1}{2m}\sum_{k=1}^P \frac{\partial f({\bf q})}{\partial q_k} \left \{ \bar p_+(q_{k-1}-\frac{\eta_{k-1}}{2},q_k+\frac{\eta_k}{2};\epsilon)   \right .  \nonumber \\ 
&& \left . \hspace{.8in}+\bar p_-(q_k-\frac{\eta_k}{2},q_{k+1}+\frac{\eta_{k+1}}{2};\epsilon) \right \} \ . \label{eq:dc_sst-10+2}
\een
The above expression, or Eq. (\ref{eq:dc_sst-10+1}), is similar to Eq. (38) of Ref. \onlinecite{hele-jcp138}, which serves as the key step in HA's suggestion\cite{hele-jcp138} that their result constitutes a derivation of  RPMD-TST\cite{craig-jcp122,craig-jcp123} However, there is an important difference, which is clarified below. The difference lies in the fact that Eq. (38) of Ref.  \onlinecite{hele-jcp138} amounts to replacing $\bar p_+$/$\bar p_-$ in Eq. (\ref{eq:dc_sst-10+2}) with $p_k$.    In other words, HA replace\cite{hele-jcp138} the average imaginary time momenta defined along the imaginary time paths at $t=0$ with the real time momenta defined by Eq. (\ref{eq:zk_pk}), which results from the short time classical-like approximation for the real time propagator.  However, the equations of motion for the real time momenta, $p_k$'s, are as yet undefined.  In fact, these simply serve as dummy integrands.  Thus, $\bar p_+$/$\bar p_-$ and $p_k$'s correspond to two distinctively different sets of variables, and the interchange between them is not justified unless stated explicitly as being an approximation, which HA did not appear to do.   

The procedure noted above appears to have resulted from a mixing of the path integral and operator formulations\cite{dirac-qm,sakurai-qm} of quantum mechanics.   For example, the ring polymer flux operator $\hat F$ defined by Eq. (32) of Ref. \onlinecite{hele-jcp138} has its meaning only within the specific convention prescribed by HA in their work.\cite{hele-jcp138} Adhering to such a convention\cite{hele-jcp138} makes their Eq. (31) equivalent to the $t=0_+$ limit of Eq. (\ref{eq:dc_sst-1}) of this work.  However, at the next stage of formulation, the momentum operator being applied to each position state is replaced with a real time value of momentum, which is not allowed quantum mechanically.    Thus, the follow-up steps of using $S({\bf q},{\bf p})$ in Eq. (33) or (38) of Ref. \onlinecite{hele-jcp138} cannot be justified 	quantum mechanically. This is tantamount to assuming that position states are eigenstates of the momentum operators in the ring polymer flux operator, which is not true.\footnote{Similar incorrect application can also be identified in going from Eq. (17) to Eq. (21) of Ref. \onlinecite{hele-jcp138}.}    Although it is possible to derive an expression in which the real time momenta appear explicitly through an alternative procedure,\footnote{S. C. Althorpe, Private Communication} as detailed in Appendix C, the resulting expression is still different from Eq. (38) of Ref. \onlinecite{hele-jcp138}.

Despite the concerns over HA's formulation as noted above, it might still be possible for the correctly calculated $t=0_+$ limit of the derivative of GKSCF, which is Eq. (\ref{eq:dc_sst-10+2}), to be equivalent to the RPMD-TST rate expression.  In order to check this, Eq. (\ref{eq:dc_sst-10+2}) can be evaluated further employing a normal mode transformation used by HA.\cite{hele-jcp138}  Let us introduce $\tilde p_l({\bf q})$ and $\tilde \eta_l ({\bf q})$ such that
\ben
&&\tilde p_l({\bf q})=\sum_{k=1}^P p_k T_{k,l}({\bf q}) \ , \\
&&\tilde \eta_l({\bf q})=\sum_{k=1}^P \eta_k T_{k,l} ({\bf q}) \ ,  
\een
where $T_{k,0}({\bf q})=B({\bf q})^{-1/2}\partial f({\bf q})/\partial q_k$ with $B({\bf q})=\sum_{k=1}^P (\partial f({\bf q})/\partial q_k)^2$.  Other components of $T_{k,l}({\bf q})$ can be determined such that $\sum_{k} T_{k,l}({\bf q}) T_{k,l'}({\bf q})=\delta_{ll'}$.  Then, Eq. (\ref{eq:dc_sst-10+2}) can be expressed as
\ben
&&k_{HA}Z_a= \frac{1}{(2\pi \hbar)^P}\int dq_{_1}\cdots \int dq_{_P}\nonumber \\
&&\hspace{.2in}\times \int d \tilde \eta_{_0}({\bf q})\cdots \int d \tilde \eta_{_{P-1}} ({\bf q}) \int \tilde dp_0 ({\bf q})\cdots \int d\tilde p_{_{P-1}} ({\bf q})\nonumber \\
&&\hspace{.2in}\times  \rho({\bf q},\boldeta) \exp (i{\bf \tilde p} \cdot \tilde \boldeta /\hbar) \Theta \left ({\bf p}\cdot \nabla f({\bf q}) \right) \delta (f ({\bf q})-d) \nonumber \\
&& \hspace{.2in} \times  \frac{1}{2m}\sum_{k=1}^P \frac{\partial f({\bf q})}{\partial q_k} \left \{ \bar p_+(q_{k-1}-\frac{\eta_{k-1}}{2},q_k+\frac{\eta_k}{2};\epsilon)   \right .  \nonumber \\ 
&& \left . \hspace{1in}+\bar p_-(q_k-\frac{\eta_k}{2},q_{k+1}+\frac{\eta_{k+1}}{2};\epsilon) \right \} \ . \label{eq:dc_sst-10+3}
\een
Integration over $\tilde \eta_k({\bf q})$ for $k\neq 0$ in the above expression can be performed easily because the integral over $\tilde p_k ({\bf q})$ amounts to a Fourier integral expression for the delta function.  Thus,
\ben
&&k_{HA}Z_a= \frac{1}{2\pi \hbar}\int dq_{_1}\cdots \int dq_{_P}\nonumber \\
&&\hspace{.2in}\times \int d \tilde \eta_{_0}({\bf q})\int \tilde dp_0({\bf q})\ \rho({\bf q},\boldeta_0) \exp (i \tilde p_0({\bf q}) \tilde \eta_0 ({\bf q}) /\hbar) \nonumber \\
&&\hspace{.2in}\times   \Theta \left (\tilde p_0 ({\bf q})\right) \delta (f ({\bf q})-d)\nonumber \\
&&\hspace{.2in} \times \frac{1}{2m}\sum_{k=1}^P \frac{\partial f({\bf q})}{\partial q_k} \left \{ \bar p_+(q_{k-1}-\frac{\eta_{0,k-1}}{2},q_k+\frac{\eta_{0,k}}{2};\epsilon)   \right .  \nonumber \\ 
&& \left . \hspace{.8in}+\bar p_-(q_k-\frac{\eta_{0,k}}{2},q_{k+1}+\frac{\eta_{0,k+1}}{2};\epsilon) \right \} \ , \label{eq:dc_sst-10+4}
\een
where $\boldeta_0=(T_{1,0},\cdots, T_{P,0})\tilde \eta_0({\bf q})$.
As detailed in Appendix D, $\bar p_+$ and $\bar p_-$ defined by Eqs. (\ref{eq:p+}) and (\ref{eq:p-}) can be calculated explicitly, resulting in Eqs. (\ref{eq:p+a1}) and (\ref{eq:p-a1}).  Employing these expressions and also performing explicit integration over $\tilde p_0({\bf q})$, under the assumption that the integrand vanishes in the limit of $\tilde p_0({\bf q})\rightarrow \infty$, we obtain the following expression: 
\ben
&&k_{HA}Z_a= \frac{1}{4\pi \hbar \epsilon}\int dq_{_1}\cdots \int dq_{_P}\nonumber \\
&&\hspace{.2in}\times \int d \tilde \eta_{_0}({\bf q})\rho({\bf q},\boldeta_0)\delta (f ({\bf q})-d)\nonumber \\
&&\hspace{.2in}\times \frac{1}{\tilde \eta_0({\bf q})}\sum_{k=1}^P \frac{\partial f({\bf q})}{\partial q_k} \Big \{ \frac{\eta_{0,k-1}({\bf q})+2\eta_{0,k}({\bf q})+\eta_{0,k+1}({\bf q})}{2}\nonumber \\ 
&&\hspace{1.4in} -q_{k-1}+q_{k+1} \Big \} \ , \label{eq:dc_sst-10+5}
\een
where $\eta_{0,k}({\bf q})=T_{k,0}\tilde \eta_0({\bf q})$. Taking the average of the above integral with an equivalent one resulting from the following variable transformation $(q_1,\cdots,q_{_P})\rightarrow (q_{_P},\cdots,q_1)$ and $\tilde \eta_0 ({\bf q}) \rightarrow -\tilde \eta_0 ({\bf q})$, we then obtain
\ben
&&k_{HA}Z_a= \frac{1}{4\pi \hbar \epsilon}\int dq_{_1}\cdots \int dq_{_P}\nonumber  \\ 
&&\hspace{.2in}\times \int d \tilde \eta_{_0}({\bf q})\rho({\bf q},\boldeta_0)\delta (f ({\bf q})-d)\nonumber \\
&&\hspace{.2in}\times \sum_{k=1}^P \frac{\partial f({\bf q})}{\partial q_k} \frac{T_{k-1,0}+2T_{k,0}+T_{k+1,0}}{2}  \ .\label{eq:dc_sst-10+6}
\een

The integration over $\tilde \eta_0({\bf q})$ in the above expression can be done explicitly following the same procedure used by HA.\cite{hele-jcp138}  Although the resulting expression is finite, which confirms the mathematical motivation behind defining the dividing surfaces of flux and side functions to coincide,  the resulting expression is clearly different from the RPMD-TST rate expression.\cite{craig-jcp122,craig-jcp123}  This also can be seen easily from the fact that Eq. (\ref{eq:dc_sst-10+6}) does not involve any real time momentum variables. 

In order to demonstrate the physical implication of the corrected HA-QTST rate expression derived above, let us consider the following case of centroid dividing surface:  
\be f({\bf q})=\frac{1}{P}\sum_{k=1}^Pq_k  \ .\ee
For this case, $\partial f/\partial q_k=1/P$, $B=1/P$, $T_{k,0}=1/\sqrt{P}$, and
\be
\sum_{k=1}^P \frac{\partial f({\bf q})}{\partial q_k} \frac{T_{k-1,0}+2T_{k,0}+T_{k+1,0}}{2}=\frac{2}{\sqrt{P}} \ .
\ee
It is also straightforward to show that 
\be 
\rho({\bf q},{\bf \eta_0})\approx \exp \Big\{ - \frac{mP}{2\hbar^2} \tilde \eta_0^2 \Big \} \rho({\bf q},0) \ .
\ee 
Inserting the above expressions into Eq. (\ref{eq:dc_sst-10+6}) and performing Gaussian integration over $\tilde \eta_0$, we find that 
\be
k_{HA} Z_a\approx \frac{1}{2\pi \hbar \beta} \rho_c(d) \ , 
\ee
where 
\be
\rho_c(d)=\sqrt{\frac{2\pi \beta \hbar^2}{m}} \int dq_1 \cdots \int dq_{_P}  \rho({\bf q},0) \delta (q_0-d) \ . 
 \ee 
 It is interesting to note that the above rate expression is exactly equal to the PI-QTST rate expression above the crossover temperature.\cite{voth-jcp91,voth-jpc97} However, this is not surprising considering that HA's formulation relies on a purely classical nature of the short real time dynamics, only within which the definition of $t=0_+$ limit makes sense.  Their definition of free energy space of the reactant, for the case of centroid dividing surface, is also the same as that for PI-QTST.\cite{voth-jcp91,voth-jpc97}  Thus, our finding above regarding the correct $t=0_+$ limit of the time derivative of the GKSCF implies that the net effect of having the flux and side dividing surfaces coincide and taking the exact $t=0_+$ limit is making classical approximation for the dynamical factor as implied in the above crossover temperature PI-QTST rate expression.\cite{voth-jcp91,voth-jpc97}

In a follow-up paper,\cite{althorpe-jcp139} Althorpe and Hele also suggest that HA-QTST is exact in a certain limit of no-recrossing.  However, their conclusion is based on the fact that any effect of a non-ideal initial condition disappears in the steady state limit.  In addition, their no-recrossing condition, which is defined as the limit where their  generalized flux-side time correlation function becomes independent of time, does not appear to naturally arise from quantum dynamical considerations.  Thus, the physical basis for the condition of no-recrossing seems unclear and difficult to verify by independent means.     

\section{Real time dynamics of the imaginary time path integral}
 Recently, we developed a formalism of RDIP,\cite{jang-cmd-rpmd,jang-jcp140} which allows calculation of a general class of nonlinear Kubo-transformed time correlation functions and thus can be used to evaluate the rigorous rate expression based on the linear response theory, Eq. (\ref{eq:kt-sf}).   
With the path integral representation of the canonical density operator given by Eq. (\ref{eq:e_beta_s}), it is straightforward to show that Eq. (\ref{eq:kt-sf}) can be expressed as
\be
k(t)= \frac{1}{Z_a}\int d{\bf q}\int d{\bf p}\ U({\bf q},{\bf p})h_{a,0} ({\bf q}) F_{0}(t;{\bf q},{\bf p}) \label{eq:cabt-1} \ , 
\ee
where $U({\bf q},{\bf p})$ is defined by Eq. (\ref{eq:u_def}) and
\ben
&&h_{a,0}({\bf q})=\frac{1}{P}\sum_{k=1}^P h_a (q_k) \ ,\\
&&F_{0}(t;{\bf q}, {\bf p}) =Tr\{\hat S(t;{\bf q}, \bar {\bf p}) \hat F\}\ ,  \label{eq:bqt_def}
\een
with  $\hat S(t;{\bf q},\bar {\bf p})\equiv e^{-i\hat Ht/\hbar}\hat S({\bf q},\bar {\bf p})e^{i\hat H t/\hbar}$, the time dependent version of $\hat S({\bf q},\bar {\bf p})$ defined by Eq. (\ref{eq:s_def}).   Note that $\bar {\bf p}$ is the midpoint momentum vector as defined below Eq. (\ref{eq:e_beta_s}).
At time $t=0$, 
\ben 
&&F_0(0;{\bf q}, {\bf p})=F_0({\bf q},{\bf p})=Tr \left \{ \hat S ({\bf q}, \bar {\bf p}) \hat F \right\} \nonumber \\
&&=\frac{1}{P}\sum_{k=1}^P \int d\eta J(q_k, \bar p_k;\eta) \langle q_k-\frac{\eta}{2}|\hat F|q_k+\frac{\eta}{2}\rangle \ . \label{eq:f0-t0}
\een
As detailed in Appendix E, the above expression is equivalent to the following classical-like expression: 
\be
F_0({\bf q},{\bf p})=\frac{1}{P} \sum_{k=1}^P \frac{\bar p_k}{m} h_b'(q_k) =\frac{1}{P}\sum_{k=1}^P \frac{\bar p_k}{m} \delta (q_k-d)\ ,  \label{eq:f0-exp}
\ee
where the second equality results from our starting assumption that $h_b(q_k)=\Theta(q_k-d)$.

Now, let us introduce ${\bf q}_{cl}(t)$ and ${\bf p}_{cl}(t)$, the path vectors consisting of $p_{k,cl}(t)$'s and $q_{k,cl}(t)$'s, each evolving classically from $\bar p_k$ and $q_k$.
Then, Eq. (\ref{eq:f0-exp}) can be expressed as
\be
F_0({\bf q},{\bf p})=\left . \frac{d}{dt} h_{b,0}({\bf q}_{cl}(t))\right |_{t=0}\ .
\ee
This motivates the following approximation: 
\ben 
F_0(t;{\bf q}, {\bf p})&\approx& F_0({\bf q}_{cl}(t),{\bf p}_{cl}(t)) \nonumber \\
&=&\frac{d}{dt} h_{b,0}({\bf q}_{cl}(t)) \ .
\een
With the above approximation, the rate expression of Eq. (\ref{eq:cabt-1}) can be expressed as
\ben
k(t)&\approx& \frac{1}{Z_a}\int d{\bf q}\int d{\bf p}\ U({\bf q},{\bf p})h_{a,0} ({\bf q}) F_0({\bf q}_{cl}(t),{\bf p}_{cl}(t)) \nonumber \\
&=&\frac{1}{Z_a}\int d{\bf q}\int d{\bf p}\ U({\bf q},{\bf p})h_{a,0} ({\bf q}) \frac{d}{dt} h_{b,0}({\bf q}_{cl}(t)) \ .  \nonumber \\ \label{eq:kt-pqst1}
\een
Evaluation of the above expression does not require quantum dynamical time evolution, and is thus feasible. 
For the case where $P=1$, the $t=0_+$ limit of this expression is equivalent to the classical TST.  However, for the quantum case, it is possible to show that the $t=0_+$ limit of the above expression is of order $1/P$, which vanishes in the exact path integral limit.  Detailed proofs are provided in Appendix F.   Thus, Eq. (\ref{eq:kt-pqst1}) satisfies the exact property that the $t=0_+$ limit vanishes, while possibly being amenable for practical calculations.   This points to the possibility of developing a path integral expression for the QTST based on the linear response theory which also reproduces the correct behavior in the $t=0_+$ limit. Previous tests\cite{jang-jcp140} of the RDIP method for harmonic oscillator was confirmed to be exact even for nonlinear operators.  Therefore, the result of the above rate expression is expected to be exact for the case of a quadratic barrier.

\section{Concluding Remarks}
The very concept of the TST relies on the principles of classical mechanics. It either assumes that instantaneous back-to-back measurements of population and flux are possible\cite{chandler-jsp42} or that completely deterministic trajectories\cite{pechukas-r1} can be defined in phase space.  In the quantum regime, these assumptions become invalid.  Therefore, it seems likely that no quantum rate theory retaining all the properties of classical TST can be found.  Even with this fundamental limitation, it may still be possible to define a quantum rate as long as the measurement of rate is defined so as to be   consistent with the history\cite{omnes-rmp64}  of quantum preparation.  Thus, a formulation of quantum rate theory approaching TST in the classical limit, which serves as the definition of QTST here, is suggested to be possible, although it does not necessarily have to be unique. 

Two popular frameworks for defining a quantum rate have been the scattering formulation\cite{miller-jcp79} and the linear response theory,\cite{yamamoto-jcp33} with the former favored in the field of gas phase dynamics and the latter in the field of condensed phase dynamics.   In both cases, a physically meaningful QTST can be defined  in the steady state limit where the reactant and product population states have completely decohered due to spatial separation or action of environments.   Only in the classical limit where the decoherence time becomes effectively zero, can such a steady state limit be replaced with the $t=0_+$ limit.  In this sense, the suggestion by HA that a true QTST can be defined as the exact $t=0_+$ limit of a quantum time correlation function and that it also reproduces the rate expression of RPMD-TST, which is approximate in nature even in the $t=0_+$ limit, is unclear.  Such a suggestion also conflicts with a significant body of earlier works on how QTST might be defined, and has therefore motivated us to carry out a detailed analysis of HA-QTST.   

In this paper, we have identified two significant concerns with the development and assumptions of HA-QTST.  First, we note that the GKSCF introduced by HA as the starting point of their QTST is not related to a time correlation function for the observable rate constant\cite{yamamoto-jcp33,voth-jpc93,stuchebrukhov-jcp95} from the linear response theory.\cite{kubo-jpsj-12,kubo-jpsj-12-2}  This is an important issue because the rate is defined as the $t=0_+$ limit of its time derivative, which should be affected significantly by the nature of the initial expression and the initial physical conditions.  Second, we identified an incorrect application of a quantum mechanical identity, namely the replacement of average imaginary time momenta along the imaginary time paths at $t=0$ with the real time momenta variables under the classical approximation at $t=0_+$.  The two are independent variables, and such a replacement is not strictly allowed quantum mechanically.  When the $t=0_+$ limit is evaluated employing a corrected expression, the resulting $t=0_+$ limit turns out to be the same as the PI-QTST rate expression above the crossover temperature\cite{voth-jcp91,voth-jpc97} for the case of centroid dividing surface.  This confirms the equivalence of assumptions of the two approaches.  In other words, the HA's formulation amounts to making classical approximation for the quantum dynamical factor and does not provide an exact quantum mechanical derivation of the RPMD-TST rate expression.  This conclusion is also consistent with our previous analysis based on the RDIP formulation\cite{jang-cmd-rpmd,jang-jcp140} showing that RPMD differs from the exact quantum dynamics even for harmonic oscillators at zero time.  

Finally, in Sec. IV,  we have presented an alternative path integral  approach for evaluating the quantum rate expression based on the exact linear response theory quantum rate expression\cite{yamamoto-jcp33,voth-jpc93,stuchebrukhov-jcp95} and by employing the formalism of  RDIP.\cite{jang-jcp140}   We have shown that the $t=0_+$ limit of this rate expression becomes zero in the quantum regime, which reproduces the known quantum mechanical behavior in the $t=0_+$ limit.  Application of this formalism to the quadratic barrier model is expected to be exact considering its result  for harmonic oscillator system.\cite{jang-jcp140}    Whether this alternative approach will also be amenable to practical calculations of general anharmonic systems will be a focus of future research.

\acknowledgments
SJ acknowledges the support for this research from  the National Science Foundation  (CHE-1362926), the Office of Basic Energy Sciences, Department of Energy (DE-SC0001393),  and the Camille Dreyfus Teacher Scholar Award.  GAV acknowledges the support of the National Science Foundation (NSF) through grant CHE-1465248.

\appendix 
\section{Review of the linear response theory formulation of quantum rate theory}
The rate expression based on the quantum linear response theory\cite{kubo-jpsj-12,kubo-jpsj-12-2} is well established,\cite{yamamoto-jcp33,voth-jpc93,stuchebrukhov-jcp95} and we here provide a brief review of the derivation for the sake of completeness.
Consider a perturbed Hamiltonian, $\hat H_\gamma =\hat H - \gamma \hat h_a$, where $\gamma$ is a small parameter.  Given that this perturbation has been imposed for a long enough time (for $t<0$), the ensemble can be represented by the following (unnormalized) canonical density operator:
\be
\hat \rho_\gamma =\exp\left \{-\beta (\hat H-\gamma \hat h_a)\right \}\ .
\ee 
Up to the first order of $\gamma$, this can be approximated as 
\be
\hat \rho_\gamma \approx e^{-\beta \hat H}+\gamma \int_0^\beta d\lambda \ e^{-(\beta-\lambda) \hat H} \hat h_a e^{-\lambda \hat H}\ .
\ee
Assume that the perturbation of $-\gamma \hat h_a$ disappears for $t\geq0$.  Then, the population of the product ($b$) within the linear response theory can be approximated as 
\ben 
&&p_b(t)\approx \frac{1}{Z+\gamma Z_a} \Big (Z_b \nonumber \\
&&+\gamma \int_0^\beta d\lambda Tr\{ e^{-(\beta-\lambda)\hat H} \hat h_a e^{-\lambda \hat H} e^{i\hat H t/\hbar}\hat h_b e^{-i\hat H t/\hbar}\} \Big ) \ ,\nonumber \\
\een  
where $Z_a=Tr\{e^{-\beta \hat H} \hat h_a\}$ and $Z_b=Tr\{e^{-\beta \hat H} \hat h_b\}$.  Expanding the denominator up to the first order and introducing the thermal equilibrium (with respect to the unperturbed Hamiltonian), $\langle \hat h_b\rangle = Z_b/Z=p_{b,eq}$, 
\be
p_b(t)\approx \frac{Z_b}{Z}\left (1+\frac{\gamma}{Z_b}\int_0^\beta d\lambda\ Tr \left \{ e^{-(\beta-\lambda) \hat H}\hat h_a e^{-\lambda\hat H} \delta \hat h_b(t)\right\}\right)\ , 
\ee
where 
\be
\delta \hat h_b(t)=e^{i\hat H t/\hbar} (\hat h_b-\langle \hat h_b\rangle )e^{-i\hat Ht/\hbar}\ .
\ee
Therefore, 
\ben
&&\delta p_b(t)=p_b(t)-p_{b,eq}\nonumber \\
&&=\frac{\gamma}{Z}\int_0^\beta d\lambda Tr\left \{e^{-(\beta-\lambda) \hat H} h_a e^{-\lambda\hat H} \delta \hat h_b(t)\right\}\ . \label{eq:delta-pbt}
\een
Taking the time derivative of this, 
\be
\frac{d}{dt} \delta p_b(t)=\frac{\gamma}{Z}\int_0^\beta d\lambda\ Tr\left \{e^{-(\beta-\lambda) \hat H}\hat h_a e^{-\lambda\hat H} \hat F(t)\right\} \ ,\label{eq:rate-1}
\ee
where $\hat F(t)$ is defined by Eq. (\ref{eq:Ft_def}).

On the other hand, given that population changes follow rate behavior near the chemical equilibrium, 
\ben
\frac{d}{dt}\delta p_b(t)&=&k_f p_a(t)-k_b p_b(t) \nonumber \\
&=&k_f(\delta p_a(t)+p_{a,eq})-k_b (-\delta p_a(t)+p_{b,eq}) \nonumber \\
&=&(k_f+k_b)\delta p_a(t)=k_f\frac{Z}{Z_b} \delta p_a(t)\ ,
\een
where $k_f$ is the forward rate and $k_b$ is the backward rate, $p_{a,eq}=Z_a/Z$, and the detailed balance condition of $k_f p_{a,eq}=k_b p_{b,eq}$ has been used. 
Under the assumption that there exists a time $t_p$ in the plateau region longer than the transient relaxation time but much smaller than the reaction time, 
\be
\left . \frac{d}{dt}\delta p_b(t)\right |_{t_p}\approx k_f\frac{Z}{Z_b}\delta p_a(0) \ .\label{eq:rate-2}
\ee  
Equating Eqs. (\ref{eq:rate-1}) and (\ref{eq:rate-2}), we obtain 
\be 
k_f\approx \frac{\gamma Z_b}{\delta p_a(0) Z^2 } \int_0^\beta d\lambda\ Tr\left \{e^{-(\beta-\lambda) \hat H}\hat h_a e^{-\lambda\hat H} \hat F(t_p)\right\}\ . \label{eq:k_f-1}
\ee
While $\delta p_a(0)$ can be evaluated directly from Eq. (\ref{eq:delta-pbt}), for the case where the barrier is high enough and the population is dominated by  those near the bottoms of the reactant and product wells, $\delta p_a(0)$ can be determined by using the detailed balance condition for the perturbed Hamiltonian.\cite{voth-jpc93,stuchebrukhov-jcp95}  In other words, for the perturbed Hamiltonian,  $Z_a^\gamma \approx Z_a(1+\beta\gamma)$ whereas $Z_b^\gamma\approx Z_b$.  Therefore, 
\be
p_a(0)\approx \frac{Z_a(1+\beta \gamma)}{Z+Z_a\beta\gamma}\ .
\ee
Then, 
\be
\delta p_a(0)=p_a(0)-p_{a,eq}= \frac{Z_a(1+\beta \gamma)}{Z+Z_a\beta\gamma} -\frac{Z_a}{Z}\approx \frac{\beta\gamma Z_aZ_b}{Z^2} \ ,
\ee 
where approximation up to the first order of $\gamma$ has been made in obtaining the second equality.  Plugging this expression into Eq. (\ref{eq:k_f-1}), we obtain  
\be
k_f\approx \frac{1}{Z_a} \frac{1}{\beta} \int_0^\beta d\lambda\ Tr \{ e^{-(\beta-\lambda)\hat H} \hat h_a e^{-\lambda \hat H} \hat F(t_p)\} \ . \label{eq:ktp-sf}
\ee
Equation (\ref{eq:kt-sf}) is a general time dependent version of this expression, from which the plateau behavior can be examined directly.

\section{Path integral representation in the complex time domain}
Consider the following expression for the quantum partition function defined in the complex time domain:
\ben
&&Z=Tr\left \{e^{-\beta \hat H}\right\} = Tr \left \{ e^{-\epsilon \hat H} e^{it \hat H/\hbar}e^{-it\hat H/\hbar} \cdots \right . \nonumber \\
&&\hspace{1.2in} \left . \times e^{-\epsilon \hat H} e^{it \hat H/\hbar}e^{-it\hat H/\hbar}\right\} \ .
\een
In the above expression, inserting $\hat 1=\int dq_i' |q_i'\rangle \langle q_i'|$ after each $e^{-\epsilon \hat H}$, $\hat 1=\int dz_i |z_i\rangle \langle z_i|$ after each $e^{it\hat H/\hbar}$, and $\hat 1=\int dq_i''|q_i''\rangle \langle q_i''|$ after each  $e^{-it\hat H/\hbar}$, we obtain
\ben
&&Z=\int dq_{_1}'\cdots \int dq_{_P}'\int dq_{_1}''\cdots \int dq_{_P}''\int dz_1\cdots \int dz_{_P}\nonumber \\
&&\hspace{.3in}\langle q_{_P}''|e^{-\epsilon \hat H}|q_{1}'\rangle \langle q_{_1}'|e^{it\hat H/\hbar}|z_1\rangle\langle z_1|e^{-it\hat H/\hbar}|q_{1}''\rangle \nonumber \\
&&\hspace{.2in}\times \langle q_{_1}''|e^{-\epsilon \hat H}|q_{2}'\rangle \langle q_{_2}'|e^{it\hat H/\hbar}|z_2\rangle\langle z_2|e^{-it\hat H/\hbar}|q_{2}''\rangle \nonumber \\
&&\hspace{.2in}\times \hspace{1in} \cdots \nonumber \\
&&\hspace{.2in}\times\langle q_{_{P-1}}''|e^{-\epsilon \hat H}|q_{_P}'\rangle \langle q_{_P}'|e^{it\hat H/\hbar}|z_{_P}\rangle\langle z_{_P}|e^{-it\hat H/\hbar}|q_{_P}''\rangle \ .\nonumber \\ \label{eq:z_path}
\een
Let us introduce
\ben
&&q_i=\frac{q_i'+q_i''}{2}\ , \label{eq:qi_def} \\
&&\eta_i=(q_i'-q_i'') \ . \label{eq:etai_def}
\een
Then, Eq. (\ref{eq:z_path}) can be expressed as follows: 
\ben
&&Z=\int dq_{_1}\cdots \int dq_{_P}\int d\eta_{_1}\cdots \int d\eta_{_P}\nonumber \\
&&\hspace{.2in}\times \int dz_1\cdots \int dz_{_P} \ \rho({\bf q},\boldeta){\mathcal G}({\bf q}, \boldeta,{\bf z};t) \label{eq:zt-cyclic}
\een
where
\ben
&&\rho({\bf q},\boldeta)=\prod_{k=1}^P\langle q_{_k}-\frac{\eta_{_k}}{2}|e^{-\epsilon \hat H}|q_{_{k+1}}+\frac{\eta_{_{k+1}}}{2} \rangle \ , \label{eq:def-rho}  \\
&&{\mathcal G}({\bf q}, \boldeta,{\bf z};t)=\nonumber \\
&&\prod_{k=1}^P \langle q_{_k}+\frac{\eta_k}{2}|e^{it\hat H/\hbar}|z_k\rangle  \langle z_k|e^{-it\hat H/\hbar}|q_{k}-\frac{\eta_k}{2}\rangle \ . \label{eq:def-g}
\een
Note that $\int dz_1\cdots \int dz_{_P} {\mathcal G}({\bf q}, \boldeta,{\bf z};t) =\delta (\eta_1)\cdots \delta (\eta_{_P})$.

\section{Evaluation of Eq. (\ref{eq:dc_sst})}
The time derivative of ${\mathcal G}({\bf q}, \boldeta, {\bf z};t)$ in Eq. (\ref{eq:dc_sst}) can be expressed as
\ben
&&\frac{d}{dt}{\mathcal G}({\bf q}, \boldeta,{\bf z};t)\nonumber \\
&&=\sum_{k=1}^P\frac{i}{\hbar}\Big \{\langle q_{k}+\frac{\eta_k}{2}|\hat H e^{it\hat H/\hbar}|z_k\rangle  \langle z_k|e^{-it\hat H/\hbar}|q_{k}-\frac{\eta_k}{2}\rangle \nonumber \\
&& \hspace{.2in}- \langle q_{k}+\frac{\eta_k}{2}|e^{it\hat H/\hbar}|z_k\rangle  \langle z_k|e^{-it\hat H/\hbar}\hat H |q_{k}-\frac{\eta_k}{2}\rangle \Big\}\nonumber \\
&& \hspace{.2in}\times \prod_{j\neq k}^P\langle q_{j}+\frac{\eta_{j}}{2}|e^{it\hat H/\hbar}|z_{j}\rangle \langle z_{j}|e^{-it\hat H/\hbar}|q_{j}-\frac{\eta_{j}}{2}\rangle \nonumber \\
&=&\sum_{k=1}^P \frac{i}{\hbar}\Big \{ \left (-\frac{\hbar^2}{2m} \left (\frac{\partial^2}{\partial x_k^2} -  \frac{\partial^2}{\partial y_k^2} \right ) +V(x_k)-V(y_k)\right) \nonumber \\
&&\hspace{.5in} \times \langle x_k|e^{it\hat H/\hbar}|z_k\rangle  \langle z_k|e^{-it\hat H/\hbar}|y_{k}\rangle  \nonumber \\ 
&&\hspace{.5in} \times \prod_{j\neq k}^P\langle x_j|e^{it\hat H/\hbar}|z_{j}\rangle \langle z_{j}|e^{-it\hat H/\hbar}|y_j\rangle \ , \label{eq:dgdt}
\een
where $x_k=q_k+\eta_k/2$ and $y_k=q_k-\eta_k/2$.
Then, Eq. (\ref{eq:dc_sst}) can be calculated by employing Eq. (\ref{eq:dgdt}), performing partial integration for the terms involving $\partial^2/\partial x_k^2$ and $\partial^2/\partial y_k^2$, and utilizing the fact that 
\ben 
&&\rho({\bf q},\boldeta)=\rho(\frac{{\bf x}+{\bf y}}{2},{\bf x-y})=\langle y_{_P}|e^{-\epsilon \hat H}|x_1\rangle \nonumber \\
&&\hspace{.2in}\times \langle y_1|e^{-\epsilon \hat H}|x_2\rangle \cdots \langle y_{_{P-1}}|e^{-\epsilon \hat H}|x_{_P}\rangle\ ,  \\
&&h_b(f({\bf q}))=h_b(f(\frac{{\bf x}+{\bf y}}{2}))  \ .
\een
For example, 
\ben
&&\int d{\bf q} \int d\boldeta \int d{\bf z}~\rho({\bf q},\boldeta)  h_b(f({\bf q}))h_b(f({\bf z})) \nonumber \\
&&\hspace{1in}\times \left (\frac{\partial^2}{\partial x_k^2} \langle x_k|e^{it\hat H/\hbar}|z_k\rangle \cdots \right ) \nonumber \\
&&=\int d{\bf q} \int d\boldeta \int d{\bf z}~{\mathcal G}({\bf q}, \boldeta,{\bf z};t) h_b(f({\bf z}))  \nonumber \\
&&\hspace{.5in}\times \frac{\partial^2}{\partial x_k^2} \left (\rho(\frac{{\bf x}+{\bf y}}{2},{\bf x-y} )  h_b(f(\frac{{\bf x}+{\bf y}}{2}))\right)\ . \nonumber \\ \label{eq:c4_app}
\een
In the integrand of the above expression, 
\ben 
&&\frac{\partial^2}{\partial x_k^2} \left (\rho(\frac{{\bf x}+{\bf y}}{2},{\bf x-y} )  h_b(f(\frac{{\bf x}+{\bf y}}{2}))\right) \nonumber \\
&&\hspace{.3in}=\Big (\frac{\partial^2}{\partial x_k^2} \rho (\frac{{\bf x}+{\bf y}}{2},{\bf x-y} ) \Big ) h_b(f(\frac{{\bf x}+{\bf y}}{2})) \nonumber \\
&&\hspace{.3in}+2\Big (\frac{\partial}{\partial x_k} \rho (\frac{{\bf x}+{\bf y}}{2},{\bf x-y} ) \Big ) \Big (\frac{\partial}{\partial x_k} h_b(f(\frac{{\bf x}+{\bf y}}{2})) \Big )\ \nonumber \\
&&\hspace{.3in}+ \rho (\frac{{\bf x}+{\bf y}}{2},{\bf x-y} )\Big (\frac{\partial^2}{\partial x_k^2} h_b(f(\frac{{\bf x}+{\bf y}}{2})) \Big ) \ . \label{eq:c6}
 \een
A similar expression can be obtained for the partial derivatives with respect to $y_k$.  Note the following identities.
\ben
&&\left (-\frac{\hbar^2}{2m} \frac{\partial^2}{\partial x_k^2} +V(x_k)\right ) \rho(\frac{{\bf x}+{\bf y}}{2},{\bf x-y} ) \nonumber \\
&&= \langle y_{_p}|e^{-\epsilon \hat H}|x_1\rangle \cdots \langle y_{k-1}|e^{-\epsilon \hat H} \hat H|x_k\rangle \cdots  \langle y_{_{P-1}}|e^{-\epsilon \hat H}|x_{_P}\rangle \ . \nonumber \\ \\
&&\left (-\frac{\hbar^2}{2m} \frac{\partial^2}{\partial y_k^2} +V(y_k)\right ) \rho(\frac{{\bf x}+{\bf y}}{2},{\bf x-y} ) \nonumber \\
&&= \langle y_{_p}|e^{-\epsilon \hat H}|x_1\rangle \cdots \langle y_{k}|\hat H e^{-\epsilon \hat H} |x_{k+1}\rangle \cdots  \langle y_{_{P-1}}|e^{-\epsilon \hat H}|x_{_P}\rangle \ . \nonumber \\
\een
When summed over all $k$, the contribution of the above two terms to the integral, Eq. (\ref{eq:c4_app}), cancel out because $e^{-\epsilon \hat H}$ commutes with $\hat H$.  In addition, the term involving the second derivative of $h_b$ with respect to $x_k$ in Eq. (\ref{eq:c6}) is equal to an analogous term involving the second derivative of $h_b$ with respect to $y_k$.  Therefore, the contribution of these terms to the integral, Eq. (\ref{eq:c4_app}) vanishes as well.  
As a result, 
\ben
&&\frac{d}{dt}\tilde C_{ss}(t) = \int dq_{_1}\cdots \int dq_{_P}\int d\eta_{_1}\cdots \int d\eta_{_P}\nonumber \\
&&\times \int dz_1\cdots \int dz_{_P}  {\mathcal G}({\bf q}, \boldeta,{\bf z};t) h_b(f({\bf z}))h_b'( f(\frac{{\bf x}+{\bf y}}{2}))\nonumber \\
&& \times  \sum_{k=1}^P \left \{\left (\frac{\hbar}{im}\frac{\partial}{\partial x_k} \rho(\frac{{\bf x}+{\bf y}}{2},{\bf x-y}  ) \right) \left (\frac{\partial}{\partial x_k} f(\frac{{\bf x}+{\bf y}}{2}) \right)   \right . \nonumber \\ 
&& -\left . \left (\frac{\hbar}{im}\frac{\partial}{\partial y_k} \rho(\frac{{\bf x}+{\bf y}}{2},{\bf x-y}  ) \right) \left (\frac{\partial}{\partial y_k} f(\frac{{\bf x}+{\bf y}}{2}) \right)   \right \} \nonumber \ ,\\ \label{eq:dc_sst-a1}
\een
where 
\ben
&&\frac{\hbar}{im}\frac{\partial}{\partial x_k} \rho(\frac{{\bf x}+{\bf y}}{2},{\bf x-y})=\nonumber \\
&&\hspace{.3in}-\frac{1}{m}\langle y_{k-1}|e^{-\epsilon \hat H} \hat p |x_k\rangle \prod_{l\neq k} \langle y_{l-1}|e^{-\epsilon \hat H}|x_l\rangle\ , \\
&&\frac{\hbar}{im}\frac{\partial}{\partial y_k} \rho(\frac{{\bf x}+{\bf y}}{2},{\bf x-y})=\nonumber \\
&&\hspace{.3in}\frac{1}{m}\langle y_{k}|\hat p e^{-\epsilon \hat H} |x_{k+1}\rangle \prod_{l\neq k} \langle y_{l}|e^{-\epsilon \hat H}|x_{l+1}\rangle\ .
\een
Inserting these expressions into Eq. (\ref{eq:dc_sst-a1}), we obtain Eq. (\ref{eq:dc_sst-1}).

For the $t=0_+$ limit of Eq. (\ref{eq:dc_sst}), an alternative expression can be found through direct evaluation of $d {\mathcal G} ({\bf q}, {\bf \eta}, {\bf z};t)/dt$ in that limit. For short enough time $t$,  
\ben
&&\langle q_k+\frac{\eta_k}{2}|e^{it\hat H/\hbar} |z_k\rangle \langle z_k|e^{-it\hat H/\hbar} |q_k-\frac{\eta_k}{2}\rangle\nonumber \\
&&\hspace{.5in}\approx \frac{m}{2\pi \hbar t} \exp\left \{-\frac{im}{\hbar t} (q_k-z_k)\eta_k\right \} \ .
\een 
Inserting the above approximation into Eq. (\ref{eq:def-g}) and using the resulting expression, Eq. (\ref{eq:c_sst}) can be expressed as 
\ben
&&\tilde C_{ss}(t) = \int dq_{_1}\cdots \int dq_{_P}\int d\eta_{_1}\cdots \int d\eta_{_P} \nonumber \\
&&\hspace{.5in}\times \int dz_1\cdots \int dz_{_P}\rho({\bf q},\boldeta)\nonumber \\
&& \hspace{.5in}\times \left (\prod_{k=1}^P  \frac{m}{2\pi\hbar t} \exp \left \{-\frac{im}{\hbar t} (q_k-z_k)\eta_k\right\} \right) \nonumber \\
&& \hspace{.5in}\times h_b\left (f({\bf q})\right) h_b \left (f({\bf z})\right)   \ .  \label{eq:c_sst-app}
\een   
Now, utilizing the variable transformation of Eq. (\ref{eq:zk_pk}), which is valid as long as $t\neq 0$, this can be expressed as 
\ben
&&\tilde C_{ss}(t) = \int dq_{_1}\cdots \int dq_{_P}\int d\eta_{_1}\cdots \int d\eta_{_P}\nonumber \\
&&\hspace{.5in}\times \int dp_1\cdots \int dp_{_P} \rho({\bf q},\boldeta) \nonumber \\
&&\hspace{.5in}\times \left (\prod_{k=1}^P  \frac{1}{2\pi\hbar } \exp \left \{\frac{im}{\hbar } p_k\eta_k\right\} \right)\nonumber \\
&&\hspace{.5in}\times  h_b\left (f({\bf q})\right) h_b \left (f({\bf q}+\frac{t}{m}{\bf p})\right)   \ . \label{eq:c_sst-app}
\een   
Taking time derivative of the above expression directly,  $\tilde C_{fs}(t)=-d\tilde C_{ss}(t)/dt$ in the short time limit turns out to have the following expression: 
\ben 
&&\tilde C_{fs}(t)= -\int dq_{_1}\cdots \int dq_{_P}\int d\eta_{_1}\cdots \int d\eta_{_P}\nonumber \\
&&\hspace{.5in}\times \int dp_1\cdots \int dp_{_P} \rho({\bf q},\boldeta)\nonumber \\
&&\hspace{.5in}\times \left (\prod_{k=1}^P  \frac{1}{2\pi\hbar} \exp \left \{\frac{im}{\hbar } p_k\eta_k\right\} \right) \Theta\left (f({\bf q})-d\right) \nonumber \\
&&\hspace{.5in}\times \delta \left (f({\bf q}+\frac{t}{m}{\bf p})-d\right )\frac{{\bf p}}{m}\cdot \nabla f({\bf q})\ .
\een  
where the fact that $h_b(x)=\Theta (x-d)$ has been used.
In the limit of $t=0_+$ this becomes
\ben 
&&\tilde C_{fs}(0_+)= -\int dq_{_1}\cdots \int dq_{_P}\int d\eta_{_1}\cdots \int d\eta_{_P}\nonumber \\
&&\hspace{.5in}\times \int dp_1\cdots \int dp_{_P} \rho({\bf q},\boldeta)\nonumber \\
&&\hspace{.5in}\times \left (\prod_{k=1}^P  \frac{1}{2\pi\hbar} \exp \left \{\frac{im}{\hbar } p_k\eta_k\right\} \right) \nonumber \\
&&\hspace{.5in} \times \Theta\left (-\frac{{\bf p}}{m}\cdot \nabla f({\bf q})0_+ \right) \nonumber \\
&&\hspace{.5in}\times \delta \left (f({\bf q})-d+\frac{{\bf p}}{m}\cdot \nabla f({\bf q})0_+\right )\frac{{\bf p}}{m}\cdot \nabla f({\bf q})\ . \nonumber \\
\een   
Despite some similarity, the above expression is different from Eq. (38) of Ref. \onlinecite{hele-jcp138} and  there appears to  be no obvious way to convert one to the other.  This is in contrast to the classical case where variable transformation from $q$ and $p$ to $q'(0_+)$ and $p'(0_+)$ can be made through Liouville's theorem as detailed in Appendix F.

\section{Evaluation of Eq. (\ref{eq:dc_sst-10}) }
For the evaluation of Eq. (\ref{eq:p+}), first consider the numerator, which can be approximated as 
\ben
&&\langle x'|e^{-\epsilon \hat H} \hat p |x'' \rangle \approx\langle x'|e^{-\epsilon \hat V} e^{-\epsilon \hat T} \hat p|x''\rangle \nonumber \\
&&=e^{-\epsilon V(x')}\frac{1}{2\pi \hbar}\int_{-\infty}^{\infty} dp\  p \ e^{-\epsilon p^2/(2m) +i(x'-x'')p/\hbar} \nonumber \\
&&=\sqrt{\frac{m}{2\pi \epsilon \hbar^2}}\frac{im(x'-x'')}{\hbar\epsilon}e^{-\epsilon V(x')-m^2(x'-x'')^2/(2\hbar^2\epsilon)}\nonumber\\
&&\approx \frac{im(x'-x'')}{\hbar\epsilon} \langle x'|e^{-\epsilon \hat H} |x'' \rangle \ .
\een
Therefore, 
\be
\bar p_+(x',x'';\epsilon)\approx \frac{im(x'-x'')}{\hbar\epsilon}\ . \label{eq:p+a1}
\ee
Similarly, one can show that
\be
\bar p_-(x',x'';\epsilon)\approx \frac{im(x'-x'')}{\hbar\epsilon}\ . \label{eq:p-a1}
\ee
Thus, to the leading order, $p_+(x',x'',\epsilon)=p_-(x',x'',\epsilon)$. 
Inserting Eqs. (\ref{eq:p+a1}) and (\ref{eq:p-a1}) into Eq. (\ref{eq:dc_sst-10}), we obtain
\ben
&&\hspace{-.2in}\tilde C_{fs}(0) \approx \int dq_{_1}\cdots \int dq_{_P} \rho({\bf q},0)\Theta(f({\bf q})-d) \delta(f({\bf q})-d)\nonumber \\
&&\hspace{.5in}\times \frac{i}{2\hbar\epsilon} \sum_{k=1}^P (q_{k-1}-q_{k+1})\frac{\partial f({\bf q})}{\partial q_k} \ .   \label{eq:dc_sst-10a}
\een
In the above integration, $q_k$'s are dummy integrands.  Therefore, the result should be invariant with respect to the following replacement: $(q_1,\cdots,q_{_P})\rightarrow (q_{_P},\cdots,q_1)$.   Upon this replacement, the summation can be rearranged such that $\sum_k (q_{k-1}-q_{k+1})\partial f/\partial q_k \rightarrow -\sum_k (q_{k-1}-q_{k+1})\partial f/\partial q_k$.  Since the two terms cancel out, this proves that $\tilde C_{fs}(0)=0$ to its leading order.  The next term is of order $O(1/\sqrt{P})$, which disappears in the $P\rightarrow \infty$ limit.

\section{Real time dynamics formulation of imaginary time path integral} 
In a recent work,\cite{jang-cmd-rpmd,jang-jcp140} we have shown that the canonical density operator for standard Hamiltonian, $\hat H=\hat p^2/(2m)+V(\hat q)$, can be expressed as
 \be
e^{-\beta\hat H}= \int d{\bf q} \int d{\bf p}\ U({\bf q},{\bf p}) \hat S({\bf q}, \bar {\bf p}) \ , \label{eq:e_beta_s}
\ee
where ${\bf q}\equiv (q_1,\cdots,q_{_P})$, ${\bf p}\equiv(p_1,\cdots,p_{_P})$, and $\bar {\bf p}\equiv (\bar p_1,\cdots, \bar p_{_P})$ with the definition of $\bar p_k=(p_k+p_{k-1})/2$ and the cyclic boundary condition $p_0=p_{_P}$.  In the above expression,  $U({\bf q}, {\bf p})$  is defined by
\ben
&&U({\bf q}, {\bf p})=\left(\frac{1}{2\pi\hbar}\right)^P \prod_{k=1}^{P}\left\{e^{-\beta V( q_k)/P}e^{-\beta p_k^2/(2mP)} \right . \nonumber \\
&&\hspace{1.5in}\times \left . e^{ip_k(q_k-q_{k+1})/\hbar} \right\}   \label{eq:u_def}\ , 
\een
with $q_{_{P+1}}=q_1$, and  $\hat S({\bf q},{\bf p})$ is defined by
\be
\hat S({\bf q},\bar {\bf p})=\frac{1}{P}\sum_{k=1}^P \int d \eta \ J(q_k,\bar p_{k};\eta)|q_k+\frac{\eta}{2}\rangle\langle q_k-\frac{\eta}{2}| \ ,\label{eq:s_def}
\ee
with 
\be 
J(q,\bar p_k;\eta)= e^{-\beta D_V(q;\eta)/P} e^{\frac{i\eta}{\hbar} \bar p_k} \ , \label{eq:Jq}
\ee 
where $ D_V (q;\eta)\equiv (V(q+\eta/2)+V(q-\eta/2))/2-V(q)$.

For the flux operator defined by Eq. (\ref{eq:flux_op}), 
\ben
&&\langle q_k-\frac{\eta}{2}|\hat F|q_k+\frac{\eta}{2}\rangle \nonumber \\
&&= \int dp' \frac{p'}{2m} \left \{ \langle q_k-\frac{\eta}{2}|p'\rangle \langle p'|q_{k}+\frac{\eta}{2}\rangle h_b'(q_k+\frac{\eta}{2})\right .\nonumber \\
&&\hspace{.7in}\left . +h_b'(q_k-\frac{\eta}{2})\langle q_k-\frac{\eta}{2}|p'\rangle \langle p'|q_k+\frac{\eta}{2}\rangle \right\} \nonumber \\
&&=\left \{h_b'(q_k+\frac{\eta}{2})+h_b'(q_k-\frac{\eta}{2})\right\} \int dp' \frac{p'}{2m} \frac{1}{2\pi\hbar} e^{-ip'\eta/\hbar}\nonumber \\
&&=\left \{h_b'(q_k+\frac{\eta}{2})+h_b'(q_k-\frac{\eta}{2})\right\} \frac{i\hbar}{2m}\frac{\partial}{\partial \eta} \delta (\eta)\ .
\een
Inserting this into Eq. (\ref{eq:f0-t0}) and conducting partial integration, we find that
\ben 
&&F_0({\bf q},{\bf p})=\frac{1}{P} \sum_{k=1}^P  \left \{ \left . -\frac{i\hbar}{m} \frac{\partial}{\partial \eta} J(q_k,\bar p_k; \eta)\right |_{\eta=0} h_b'(q_k) \right . \nonumber   \\
&&\hspace{.2in}  -\frac{i\hbar}{2m} J(q_k,\bar p_k;0)\nonumber \\
&&\hspace{.2in} \times \left . \left . \left (\frac{\partial}{\partial \eta} h_b'(q_k+\frac{\eta}{2})+\frac{\partial}{\partial \eta} h_b'(q_k-\frac{\eta}{2}) \right )\right |_{\eta=0} \right \}\ . \label{eq:f0-t0-2}
\een
In the above expression, the two terms within the parenthesis of the last line, which involves derivatives of $h_b'(q_k\pm \eta/2)$, cancels out.     
Employing the expression for Eq. (\ref{eq:Jq}), it is easy to show that the derivative of $J(q_k,\bar p_k;\eta)$ with respect to $\eta$ at $\eta=0$ results in $i\bar p_k/\hbar$.  
As a result,  Eq. (\ref{eq:f0-t0-2}) simplifies to Eq. (\ref{eq:f0-exp}).

\section{$t=0_+$ limit of Eq. (\ref{eq:kt-pqst1})}
In the limit of $t\rightarrow 0_+$,  Eq.  (\ref{eq:kt-pqst1}) approaches 
\ben
k(0_+)&=&\frac{1}{Z_A}\int d{\bf q}\int d{\bf p}\ U({\bf q},{\bf p}) \left (\frac{1}{P}\sum_{k=1}^P h_a(q_k)\right) \nonumber \\
&&\hspace{.5in}\times \left (\frac{1}{P}\sum_{k=1}^P \frac{\bar{p}_{k}}{m} h_b'(q_{k,cl}(0_+))\right) \ . \label{eq:zero-time-qtst}
\een
The classical analog of this expression corresponds to the case where $P=1$ and $U({\bf q},{\bf p})$ is replaced with $e^{-\beta H}/(2\pi \hbar)$.  
Thus, 
\be
k_{cl}(0_+)= \frac{1}{Z_a}\int \int \frac{dq d p}{2\pi\hbar} e^{-\beta H(q,p)} h_a(q)\frac{p}{m} h_b'(q(0_+)) \ .\label{eq:ctst}
\ee
Let us consider the  simple case where $h_b(q)=\Theta (q-d)$. First, due to the Liouville theorem, $dqdp\ e^{-\beta H(q,p)}$ in the integrand can be replaced with $dq'dp'\ e^{-\beta H(q',p')}$ where $q'= q(0_+)$ and $p'= p(0+)$.  Then, $q=q'(0_-)$. Dropping primes, we then obtain the following expression: 
\ben
k_{cl}(0_+)&=&\frac{1}{Z_a}\int \int \frac{dq d p}{2\pi\hbar} e^{-\beta H(q,p)} h_a(q(0_-))\frac{p}{m} \delta(q-d) \nonumber \\
&=&\frac{1}{Z_a}\int \int \frac{dq d p}{2\pi\hbar} e^{-\beta H(q,p)} \Theta(p)\frac{p}{m} \delta(q-d) \nonumber \\
&=&k_{cl}^{TST}\ , \label{eq:kcl_tst}
\een
where the following relation has been used.  
\ben
&&h_a(q(0_-))\delta (q-d)=\lim_{\epsilon\rightarrow 0} (1-\Theta (-\epsilon p) )\delta (q-d)\nonumber \\
&&=\Theta (p)\delta (q-d) \ .
\een
Equation (\ref{eq:kcl_tst}) is the well-known classical TST rate expression.  

In contrast to the above classical case, the zero time limit for the quantum case, Eq. (\ref{eq:zero-time-qtst}) can be shown to be zero. 
In order to show this, the double summation in Eq. (\ref{eq:zero-time-qtst}) can be divided into two terms as follows:
\ben
&&k(0_+)=\frac{1}{Z_a}\int d{\bf q}\int d{\bf p}\ U({\bf q},{\bf p}) \nonumber \\
&&\hspace{.2in}\times \frac{1}{P^2}\left (\sum_{k=1}^P h_a(q_k)\frac{\bar{p}_{k}}{m} h_b'(q_{k,cl}(0_+) )\right  .  \nonumber \\
&&\hspace{.5in} +\left . \sum_{k=1}^P\sum_{l\neq k}^P h_a(q_k)\frac{\bar{p}_{l}}{m} h_b'(q_{l,cl}(0_+) ) \right ) \ .\label{eq:zero-time-qtst-2}
\een
For the case where $h_b(q)=\Theta(q-d)$, going through the same procedure of variable transformation and time translation as in deriving Eq. (\ref{eq:kcl_tst}), we obtain
\ben
&&k(0_+)=\frac{1}{Z_a}\int d{\bf q}\int d{\bf p}\ U({\bf q},{\bf p}) \nonumber \\
&&\hspace{.2in}\times \frac{1}{P^2}\left (\sum_{k=1}^P \Theta (\bar p_k)\frac{\bar{p}_{k}}{m} \delta(q_{k}-d) \right  .  \nonumber \\
&&\hspace{.5in} +\left . \sum_{k=1}^P\sum_{l\neq k}^P h_a (q_k)\frac{\bar{p}_{l}}{m} \delta(q_l-d ) \right )\ .\label{eq:zero-time-qtst-3}
\een
In the above expression, the first term involving single sum over $k$ is nonzero as in the classical case but its contribution is of order $(1/P)$ and vanishes in the limit of $P\rightarrow\infty$.
The second term involving double summation can be evaluated explicitly and shown to be zero due to symmetry as follows:
\ben
&&\int d{\bf q}\int d{\bf p}\ U({\bf q},{\bf p}) \frac{1}{P^2}\sum_{k=1}^P\sum_{l\neq k}^P h_a (q_k)\frac{\bar{p}_{l}}{m} \delta(q_l-d ) \nonumber \\ 
&&=\frac{i}{2\beta\hbar P} \left (\frac{2\pi m P}{\beta}\right)^{P/2}\int d{\bf q}\ e^{-\beta V_{RP} ({\bf q})/P}\nonumber \\
&&\hspace{.1in}\times \sum_{k=1}^P\sum_{l\neq k}^P h_a(q_k) (q_l-q_{l+1})(\delta(q_l-d )+\delta (q_{l+1}-d)) \nonumber \\ 
&&=\frac{i}{4\beta\hbar P} \left (\frac{2\pi m P}{\beta}\right)^{P/2}\int d{\bf q} e^{-\beta V_{RP} ({\bf q})/P} \sum_{k=1}^P\sum_{l\neq k}^P h_a (q_k)\nonumber \\ 
&&\hspace{.2in}\times (q_l-q_{l+1} +q_{l+1}-q_{l}) (\delta(q_l-d )+\delta (q_{l+1}-d)) \nonumber \\
&&=0 \ . \label{eq:zero-time-qtst-4}
\een
where
\be
V_{RP}({\bf q})=\sum_{k=1}^P\left \{ V(q_k)+\frac{mP^2}{2\beta^2\hbar^2}(q_k-q_{k+1})^2\right\} \ .
\ee
In Eq. (\ref{eq:zero-time-qtst-4}), the first equality is obtained by explicit integration over $p_k$'s, and the second equality is obtained using the fact that $V_{RP}({\bf q})$ is invariant with respect to exchange of  $q_l \rightarrow q_{l+1}$ and the reversal of the ordering of indices.

\end{document}